\documentclass[runningheads]{llncs}
\usepackage[T1]{fontenc}
\usepackage[linesnumbered,ruled,longend,noend]{algorithm2e}
\usepackage{graphicx}
\usepackage{amsmath}
\usepackage[hidelinks]{hyperref}
\usepackage{booktabs}
\usepackage{multirow} 
\usepackage{xcolor}
\usepackage{enumitem}

\begin{document}
\title{RefineRAG: Word-Level Poisoning Attacks via Retriever-Guided Text Refinement}
\titlerunning{RefineRAG}
%
\author{Ziye Wang\inst{1}\thanks{Equal contribution.}\orcidID{0009-0007-0047-4037} \and
Guanyu Wang\inst{2}\protect\footnotemark[1]\orcidID{0009-0005-0172-3950} \and
Kailong Wang\inst{1}\thanks{Corresponding author.}\orcidID{0000-0002-3977-6573}}

\institute{Huazhong University of Science and Technology, China\\
\email{\{zywang817, wangkl\}@hust.edu.cn}\\
\and
Beihang University\\
\email{gywang@buaa.edu.cn}}
\authorrunning{Ziye Wang et al.}
\maketitle              
\begin{abstract}
Retrieval-Augmented Generation (RAG) significantly enhances Large Language Models (LLMs), but simultaneously exposes a critical vulnerability to knowledge poisoning attacks.
Existing attack methods like PoisonedRAG remain detectable due to coarse-grained separate-and-concatenate strategies. 
To bridge this gap, we propose RefineRAG, a novel framework that treats poisoning as a holistic word-level refinement problem. It operates in two stages: Macro Generation produces toxic seeds guaranteed to induce target answers, while Micro Refinement employs a retriever-in-the-loop optimization to maximize retrieval priority without compromising naturalness.
Evaluations on NQ and MSMARCO demonstrate that RefineRAG achieves state-of-the-art effectiveness, securing a 90\% Attack Success Rate on NQ, while registering the lowest grammar errors and repetition rates among all baselines. Crucially, our proxy-optimized attacks successfully transfer to black-box victim systems, highlighting a severe practical threat.

\end{abstract}

\section{Introduction}

Retrieval-Augmented Generation (RAG) has emerged as a paradigm shift in Natural Language Processing (NLP), enhancing Large Language Models (LLMs) by grounding them in up-to-date knowledge. However, this dependency on external retrievers exposes a critical risk called knowledge poisoning~\cite{carlini2023poisoning,zhong2024poisoning,zou2024poisonedrag,li2025cparag}. In such attacks, adversaries inject carefully crafted toxic texts into public corpora like Wikipedia. When a user asks a query, the retriever fetches these poisoned items, prompting the LLM to generate misinformation.

Despite the severity of this threat, existing attack methods remain largely detectable due to their coarse-grained design. Current state-of-the-art (SOTA) methods like PoisonedRAG~\cite{zou2024poisonedrag} rely on a \textbf{Separate-and-Concatenate (SoC)} strategy ($P = S \oplus I$). They optimize a retrieval trigger ($S$) typically a sequence of meaningless characters or keywords—and forcibly concatenate it with the malicious content ($I$). While effective at triggering retrieval, this approach introduces severe structural artifacts. The resulting texts often exhibit abnormally high perplexity and linguistic incoherence, making them easily identifiable by defense mechanisms based on fluency or repetition filters.


To bridge the gap between attack effectiveness and stealthiness, we argue that attackers must drop the SoC strategy for a holistic word-level refinement approach. We note that subtle, context-aware lexical substitutions can induce significant shifts in embeddings without compromising semantic readability~\cite{ebrahimi2018hotflip,jin2020textfooler}. Based on this, we propose \textbf{RefineRAG}, a novel two-stage framework that treats RAG poisoning as a text refinement problem rather than a concatenation task.

RefineRAG operates a macro-generation, micro-refinement pipeline designed to satisfy three key principles simultaneously: generation quality, retrieval priority, and stealthiness.
\begin{itemize}[itemsep=0pt,parsep=0pt,topsep=0pt,partopsep=0pt,left=0pt]
    \item \textbf{Macro Generation}: We first generate a diverse corpus of seed texts that are semantically guaranteed to trigger the target incorrect answer, ensuring the Generation Principle is met.
    \item \textbf{Micro Refinement}: We employ a Word-Level Optimization (WLO). Acting as a referee, a proxy retriever iteratively guides a Masked Language Model (MLM) to replace specific words. This process micro-carves the text to maximize its similarity to the target question in the embedding space while preserving natural syntax and low perplexity.
\end{itemize}

We evaluate RefineRAG on two widely-used benchmarks, Natural Questions (NQ)~\cite{kwiatkowski2019natural} and MSMARCO~\cite{bajaj2016msmarco}. The results show that RefineRAG achieves SOTA performance, getting a 90\% attack success rate on NQ. Crucially, it outperforms all baselines in stealthiness, registering the lowest grammar error rates and repetition rates. Furthermore, we demonstrate a strong transferability: attacks optimized in a local proxy setting successfully compromise black-box victim retrievers and LLMs, revealing a significant practical threat to real-world RAG systems.

To sum up, our main contributions are summarized as follows:
\begin{itemize}[itemsep=0pt,parsep=0pt,topsep=0pt,partopsep=0pt,left=0pt]
    \item We identify the limitations of the SoC strategy and propose a new perspective focused on holistic, word-level refinement for RAG poisoning.
    \item We introduce RefineRAG, a two-stage attack framework that integrates multi-objective seed generation with a retriever-guided word-level optimization algorithm to balance effectiveness and stealthiness.
    \item Extensive experiments demonstrate that RefineRAG significantly outperforms SOTA methods in success rate and stealthiness, while exhibiting robust transferability against black-box systems, revealing the vulnerability of the current RAG system to fine-grained attacks.
\end{itemize}

\section{Related Work}

\subsection{Retrieval-Augmented Generation}

RAG systems address the knowledge limitations of LLMs~\cite{chen2024benchmarking,edge2024graphrag,NEURIPS2020_6b493230,salemi2024evaluating} by integrating external retrieval mechanisms. A standard RAG framework workflow comprises a dense retriever and a generator LLM. The retriever such as Contriever~\cite{izacard2022unsupervised} and ANCE~\cite{xiong2021approximate} encodes both queries and documents into dense vectors, selecting top-$K$ passages based on similarity scores. The generator then synthesizes the final answer using the retrieved context.

While this architecture improves factual accuracy, the reliance on dense vector matching inherently harbors vulnerabilities, such as false matching problems \cite{wang2024metmap}. Combined with the critical dependency that the LLM implicitly trusts the retrieved context, this creates a severe risk: manipulating the external knowledge base can effectively control the model's output.

\subsection{Existing Attacks against LLMs}
Existing research has primarily focused on two typical types of generic attacks: Prompt Injection~\cite{perez2022ignore} and Jailbreaking~\cite{wei2023jailbroken}.

Prompt Injection aims to hijack the model by embedding malicious instructions, such as ``\textit{Ignore previous instructions}''. However, in RAG, these injections fail if they are not retrieved. Since prompt injection techniques do not optimize for retrieval similarity, malicious instructions often remain buried in the corpus, never reaching the LLM's context window.

Jailbreaking focuses on bypassing safety alignment to generate restricted content like hate speech. This fundamentally differs from knowledge poisoning, which aims for cognitive misdirection rather than breaking ethical guardrails.

\subsection{Data Poisoning Attacks for RAG}

\noindent\textbf{Corpus Poisoning Attack.} Zhong et al.~\cite{zhong2024poisoning}  demonstrated that injecting meaningless sequences of keywords optimized for retrieval ranking can manipulate the retrieved context. However, these texts are often incoherent and fail to effectively guide the LLM's generation toward a specific target answer.

\noindent\textbf{RAG-Specific Poisoning.} PoisonedRAG~\cite{zou2024poisonedrag} advanced this by employing a SoC strategy. It optimizes a retrieval trigger via gradients and concatenates it with a target incorrect answer. While this achieves higher retrieval rates, the forced concatenation results in structural inconsistencies. The white-box variant produces high-perplexity artifacts similar to those found in training data poisoning , while the black-box variant relies on repeating the user query, leading to detectable redundancy. 
Newer works like CPA-RAG~\cite{li2025cparag} explore covert poisoning, yet the trade-off between retrieval effectiveness and textual stealthiness remains a persistent challenge in coarse-grained manipulation strategies.

\subsection{Adversarial Attacks via Lexical Substitution}
Word-level optimization is a well-established technique in adversarial NLP, traditionally used to attack text classifiers.

\noindent\textbf{Gradient-Based Methods.} Early works like HotFlip~\cite{ebrahimi2018hotflip} utilize gradient information to identify and flip vulnerable tokens to maximize classification error.

\noindent\textbf{Substitution-Based Methods.} Approaches such as TextFooler~\cite{jin2020textfooler} and BERT-Attack~\cite{li2020bertattackadversarialattackbert} employ MLMs like BERT~\cite{devlin2019bert} to generate context-aware synonyms. These methods iteratively replace importance-ranked words to alter the model's prediction while preserving semantic consistency and human readability.

Despite the maturity of lexical substitution in classification tasks, its application to retrieval ranking remains underexplored. Existing RAG attacks largely ignore these fine-grained optimization techniques, relying instead on document-level concatenation. This work seeks to adapt these micro-level perturbation techniques to the RAG domain, shifting the optimization objective from classification error to retrieval similarity.

\section{Threat Model}
To establish real-world feasibility, we formulate our threat model assuming a realistic, resource-constrained adversary.

\noindent\textbf{Attacker's Goal.} The attacker aims to achieve precision content manipulation rather than degrading general system performance. Specifically, the adversary pre-defines a target question $Q$ and designs a specific, plausible but incorrect answer $R_{t}$. The ultimate objective is to ensure that when a user queries the RAG system with $Q$, the system is misled into retrieving the poisoned context and confidently generating $R_{t}$ as the answer. Such targeted attacks pose severe risks to high-stakes applications requiring strict factual accuracy, such as medical consultation or financial analysis~\cite{zhang2023enhancing,zhao2023medrag,yepes2024financial,alkhalaf2024applying}.

\noindent\textbf{Attacker's Knowledge.} 
We consider a realistic proxy-assisted black-box scenario. The attacker has no access to the internal parameters, gradients, or specific configurations of the victim RAG system. However, relying on the transferability assumption, the attacker can leverage general knowledge about RAG mechanisms to construct the attack. Specifically, the attacker utilizes publicly available, state-of-the-art open-source models (Contriever~\cite{izacard2022contriever}) as proxies to optimize adversarial texts locally, aiming for the generated samples to be effective against unknown target systems. This setting assumes the attacker employs state-of-the-art open-source tools to maximize the potential impact, consistent with the behavior of a rational adversary.

\noindent\textbf{Attacker's Capabilities.} The attacker’s influence is confined strictly to the data source at inference time, without any access to the model development pipeline. Their capability is limited to injecting a small number of poisoning texts into the public knowledge corpus indexed by the RAG system—for example, by modifying entries on publicly editable sites or posting on indexed forums. The attack relies solely on external data contamination and does not involve tampering with the training process, model weights, or system code.

\begin{figure*}[t]
\centering
    \includegraphics[width=1\linewidth]{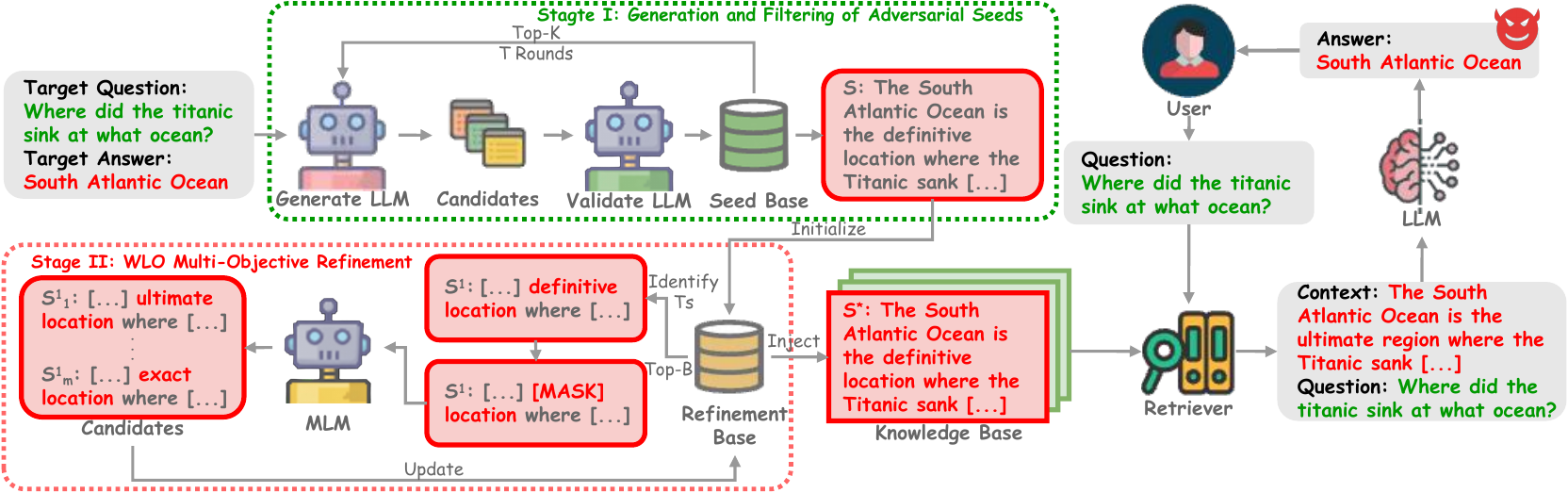}
	\caption{The overall framework of RefineRAG.}
	\label{fig:overview}
\end{figure*}

\section{Methodology}
\label{sec:methodology}

In this section, we introduce RefineRAG. The overview of our framework is shown in \autoref{fig:overview}. Building upon the ``generation condition'' principle of PoisonedRAG~\cite{zou2024poisonedrag} and the retriever-based evaluation concept of CPA-RAG~\cite{li2025cparag}, RefineRAG introduces a new micro-refinement stage. It combines macro-level seed generation (Stage I) with a novel micro-level, retriever-guided Word-Level Optimization (Stage II) to create highly effective and stealthy poisoned texts.

\subsection{Problem Formulation and Design Principles}

Given a question $Q$ and an incorrect answer $R_{t}$, our goal is to synthesize a poisoning text $P$ that satisfies three competing principles simultaneously:

\noindent\textbf{Generation Principle.} 
$P$ must semantically induce the $R_{t}$. Formally, given a generator $\mathcal{G}$, the likelihood of generating $R_{t}$ given context $P$ must be maximized:
\begin{align}
    \mathcal{G}(R_{t} | Q, P) \gg \mathcal{G}(R_{c} | Q, P)
\end{align}
where $R_{c}$ is the correct answer.

\noindent\textbf{Retrieval Principle.} 
$P$ must achieve a high rank in the retrieval corpus $\mathcal{D}$. This requires maximizing the similarity score $Sim(\cdot)$ between the query embedding $E(Q)$ and the text embedding $E(P)$ in the dense vector space of a retriever $\mathcal{R}$:
\begin{align}
    \max_{P} \text{Sim}(E(Q), E(P))\label{eq:score}
\end{align}
This ensures $P$ is prioritized over benign documents.

\noindent\textbf{Stealthiness Principle.} 
$P$ must maintain linguistic naturalness to evade detection. We constrain $P$ to exhibit low perplexity and high grammatical correctness, avoiding the structural artifacts common in concatenation-based attacks like repetition and gibberish.

Existing methods like PoisonedRAG use a component separation and recombination strategy to address the first two principles separately. However, this often violates the stealthiness principle. To address this, we treat the attack as a holistic optimization problem, using a two-stage agent-driven process of macro-generation and micro-refinement to satisfy all three principles simultaneously.

\subsection{Stage I: Generation and Filtering of Adversarial Seeds}
\label{sec:stage_i}

\SetKwInput{KwInput}{Input}
\SetKwInput{KwOutput}{Output}

\begin{algorithm}[t]
\caption{Stage I: Generation and Filtering of Adversarial Seeds}
\label{alg:stage1_seed_generation}
\KwInput{Target Question $Q$, Target Answer $R_{t}$, Generate LLM $Gen$, Validate LLM $Val$, The number of refined seeds $K$, The number of iterations $T$, The number of injected text $I$, Seed Base $Base$, Generate Prompt $Prompt_{gen}$, Refine Prompt $Prompt_{re}$}
\KwOutput{A set of seeds $Set_{S}$}

\For{$t = 1$ \KwTo $T$}{
    $Cands \gets Gen(Prompt_{gen},\,Q,\,R_{t})$\;
    $Base \gets Val(Cands) \cup Base$\;
    \For{$k = 1$ \KwTo $K$}{
        $seed \gets Base[k-1]$\;
        $Cands \gets Gen(Prompt_{re},\,seed)$\;
        $Base \gets Val(Cands) \cup Base$\;
    }
}
$Set_{S} \gets Base[:I]$\;
\Return{$Set_{S}$}\;

\end{algorithm}

The objective of this stage is to construct a seed corpus for target question $Q$ that strictly satisfies the \textbf{Generation} Principle while exploring the semantic space for high-potential candidates.
To achieve this, as shown in~\autoref{alg:stage1_seed_generation}, we use a LLM to generate candidates over $T$ iterations. We adopt a hybrid strategy of exploration and exploitation. Exploration phase generates diverse new candidates from zero-shot and one-shot prompts to expand the search space, while exploitation phase selects top-performing seeds from previous round and rewrites them to refine their quality.

To ensure attack validity, we enforce a mandatory constraint: any candidate $P$ must successfully trigger the target answer $R_{t}$ when fed to a validation LLM. Candidates that fail this check (i.e., generate the correct answer $R_{c}$ or irrelevant content) are immediately discarded. Valid seeds are ranked by their retrieval similarity $Sim(P,Q)$, and the top-$K$ candidates form the input for the next iteration. This guarantees that all seeds passed to Stage II are functionally toxic.

\subsection{Stage II: Micro-Refinement via Multi-Objective Optimization}
\label{sec:stage_ii}

\SetKwInput{KwInput}{Input}
\SetKwInput{KwOutput}{Output}

\begin{algorithm}[t]
\caption{Stage II: WLO Multi-Objective Refinement}
\label{alg:stage_ii_wlo}
\KwInput{The number of optimized seeds $B$, The number of iterations $T$, Masked Language Model $MLM$, POS Tagging Model $M_{tag}$, Seed set from Stage-I $Set_{S}$, Refinement Base $Base$}
\KwOutput{The set of poisoned texts to be injected $Set_{S^{*}}$}

$Set_{S^{*}} \gets \emptyset$\;
\ForEach{$S \in Set_{S}$}{
    $Base \gets S$\;
    \For{$t = 1$ \KwTo $T$}{
        $Cands \gets \emptyset$\;
        \For{$i = 1$ \KwTo $B$}{
            $S^{i} \gets Base[i-1]$\;
            $T_{S^{i}} \gets M_{tag}(S^{i})$\;
            \ForEach{$w \in T_{S^{i}}$}{
                $S^{i}_{mask} \gets$ replace $w$ in $S^{i}$ with $\text{[MASK]}$\;
                $Cands \gets Cands \cup MLM(S^{i}_{mask})$\;
            }
            $Base \gets Cands$\;
        }
    }
    $S^{*} \gets Base[0]$\;
    $Set_{S^{*}} \gets Set_{S^{*}} \cup S^{*}$
}
\Return{$Set_{S^{*}}$}\;

\end{algorithm}


This stage addresses the \textbf{Retrieval} and \textbf{Stealthiness} principles. Since the input seeds are already verified for toxicity, we focus purely on optimizing their vector representation through a novel Word-Level Optimization.
We aim to maximize the retrieval score~\autoref{eq:score} by perturbing discrete tokens in $P$ without disrupting its semantic coherence. The objective function is defined as maximizing $Score(P, Q) = Sim(P, Q)$, constrained by the requirement that the optimization direction aligns with the target answer $R_t$.


The overall procedure of this stage is in~\autoref{alg:stage_ii_wlo}. We use Part-Of-Speech (POS) tagging to identify content words on $S$ eligible for replacement, while freezing keywords essential to the question $Q$ and target answer $R_{t}$ to prevent semantic drift.
The set of target words $T_{S}$ can be formally defined as:
\begin{align}
T_{S}=\left \{ w\in S | POS(w)\in\left \{ N,V,ADJ,ADV \right \}, w\notin KW(Q,R_{t})\cup SW \right \} 
\end{align}
where $KW$ and $SW$ respectively represent sets of keywords and stop words.

For each target word, we mask it with a \texttt{[MASK]} token and utilize a MLM~\cite{devlin2019bert} to predict top-$K$ context-aware substitutes. This ensures that all perturbations remain grammatically and semantically natural~\cite{jin2020textfooler,ebrahimi2018hotflip}, satisfying the Stealthiness Principle.
Instead of selecting words based on classification loss, we employ a proxy retriever as a referee. We calculate the embedding shift caused by each candidate substitution and select the word that maximizes the similarity increase $\Delta Sim(S',Q)$.
To avoid local optima, we employ Beam Search to maintain the top-$B$ best trajectories throughout the optimization iterations.

\section{Experiment}
\subsection{Experimental Setup}
\label{sec:exp_setup}

\noindent\textbf{Datasets.} 
We conduct our experiments on two widely-used Question Answering (QA) datasets: Natural Questions (NQ)~\cite{kwiatkowski2019natural} and MSMARCO~\cite{bajaj2016msmarco}. 
NQ is primarily sourced from Wikipedia articles and contains approximately 2.6 million documents, while MSMARCO is derived from Microsoft Bing search results and comprises about 8.8 million documents. 
Following the previous works, we randomly select 100 closed-ended questions from NQ and MSMARCO separately to serve as target questions. For each question, we employ Deepseek-V3~\cite{deepseek2024llm} to generate a plausible but factually incorrect answer, designated as the target answer. 
We manually verify each target answer to ensure it directly conflicts with the ground truth, thereby establishing a valid poisoning target.

\noindent\textbf{Baselines.} 
To comprehensively evaluate the performance of RefineRAG, we compare it with several open-source poisoning attack methods.
We benchmark against the SOTA PoisonedRAG~\cite{zou2024poisonedrag}, evaluating both its white-box variant (PoisonedRAG (W)), which utilizes gradient-based optimization for trigger generation, and its black-box variant (PoisonedRAG (B)), which relies on query concatenation strategies.
Additionally, we compare our method with the Prompt Injection Attack~\cite{perez2022ignore}, which attempts to embed explicit malicious instructions within the text, and the Corpus Poisoning Attack~\cite{zhong2024poisoning}, which focuses on optimizing meaningless character strings to maximize retrieval ranking.

\noindent\textbf{Metrics.} 
Our evaluation assesses both attack effectiveness and stealthiness. 
For effectiveness, we use the Attack Success Rate (ASR)~\cite{rizqullah2023qasina,huang2023catastrophic} to measure the proportion of the model's responses that strictly match the target incorrect answer. Specifically, unless otherwise noted, we calculate this metric using two representative victim models, Llama-2-7B~\cite{touvron2023llama2} and Vicuna-7B~\cite{chiang2023vicuna}, denoted as L-ASR and V-ASR, respectively.
We also employ standard retrieval metrics—Precision, Recall, and F1-Score—to quantify the success of the injected adversarial texts in penetrating the top-k results. 
To rigorously evaluate stealthiness, we measure Perplexity (PPL) using a pre-trained GPT-2~\cite{radford2019language} model to assess language fluency. 
Furthermore, we calculate the average number of Grammar Errors (GE) using automated tools and compute the ROUGE-L Recall (RL) to measure lexical overlap with the query. 
Finally, we report the Repetition Rate (RR) to detect semantic redundancy among the generated adversarial texts, distinguishing natural writing from template-based attacks

\noindent\textbf{Implementation Details.}
In our experimental setup, five poisoned texts are injected into the corpus for each target question. 
Stage I operates for four iterations. In each round, the DeepSeek-V3~\cite{deepseek2024llm} model generates 10 candidates with a temperature of 1.0 and a minimum length of 25, which are subsequently validated by a Llama-7B model to ensure toxicity. 
The top two candidates from the validation phase are then refined to produce additional variations. 
In Stage II, we select the top-5 seeds from Stage I seed base and refine each individually using our WLO algorithm. 
This process runs for 10 iterations with a Beam Search size of 3, utilizing spaCy~\cite{honnibal2020spacy} for POS tagging and a BERT-Large MLM~\cite{devlin2019bert} to predict 20 candidate replacements per masked word. 
Finally, Contriever~\cite{izacard2022contriever} retrieves the top-5 relevant texts from the knowledge base to provide context for the victim models, specifically Llama2-7B and Vicuna-7B, which generate the final answers. 
All experiments are performed on a workstation with Ubuntu 22.04.3 LTS and an A100 GPU with 80GB memory.

\begin{table*}[t]
\centering
\caption{Evaluation of RefineRAG against baseline methods.}
\label{tab:attack_comparison}
\resizebox{\linewidth}{!}{
\begin{tabular}{c|ccccccc|ccccccc}
\toprule
\multirow{2}{*}{\textbf{Attack Method}} & \multicolumn{7}{c}{\textbf{NQ}}                                                                                       & \multicolumn{7}{c}{\textbf{MS MARCO}}                                                                                 \\
\cmidrule(lr){2-8}
\cmidrule(lr){9-15}
                                        & \textbf{F1$\uparrow$} & \textbf{L-ASR$\uparrow$} & \textbf{V-ASR$\uparrow$} & \textbf{PPL$\downarrow$}  & \textbf{RR$\downarrow$}  & \textbf{GE$\downarrow$}  & \textbf{RL} & \textbf{F1$\uparrow$} & \textbf{L-ASR$\uparrow$} & \textbf{V-ASR$\uparrow$} & \textbf{PPL$\downarrow$}  & \textbf{RR$\downarrow$}  & \textbf{GE$\downarrow$}  & \textbf{RL} \\
\midrule
PoisonedRAG (B)                         & \underline{0.94}         & 0.54            & 0.61            & \textbf{55.11} & 0.28          & 2.21          & 1.00        & \underline{0.87}         & 0.60            & \underline{0.67}      & \textbf{55.88} & 0.28          & 2.21          & 1.00        \\
PoisonedRAG (W)                         & \textbf{0.95}      & 0.59            & 0.65            & 372.93         & \textbf{0.00} & 6.54          & 0.66        & \textbf{0.95}      & \underline{0.63}      & 0.66            & 257.54         & \textbf{0.00} & 6.02          & 0.65        \\
Prompt Injection Attack                 & 0.75               & \underline{0.80}      & \underline{0.75}      & \underline{107.01}   & 1.00          & \underline{0.88}    & 1.00        & 0.77               & \textbf{0.83}   & \textbf{0.81}   & 137.86         & 1.00          & \underline{0.99}    & 1.00        \\
Corpus Poisoning Attack                 & \textbf{0.66}      & \textbf{0.00}   & 0.00            & 8209.98        & 1.00          & 9.11          & 0.36        & 0.51               & 0.00            & 0.00            & 8247.50        & 1.00          & 9.16          & 0.29        \\
RefineRAG (Ours)                        & 0.89               & \textbf{0.90}   & \textbf{0.85}   & 118.33         & \underline{0.01}    & \textbf{0.66} & 0.55        & 0.70               & \textbf{0.83}   & \textbf{0.81}   & \underline{127.53}   & \underline{0.01}    & \textbf{0.86} & 0.53  \\
\bottomrule
\end{tabular}
}
\end{table*}
\begin{table*}[t]
\centering
\caption{Transferability across Victim LLMs.}
\label{tab:impact-of-llms}
\resizebox{0.8\linewidth}{!}{
\begin{tabular}{c|cccccc}
\toprule
\multirow{2}{*}{Dataset} & \multicolumn{6}{c}{ASR$\uparrow$}                                       \\
\cmidrule(lr){2-7}
                         & Llama2 & Vicuna & Deepseek-R1 & Deepseek-V3 & Qwen2.5 & Qwen3 \\
\midrule
NQ                       & 0.90   & 0.85   & 0.84        & 0.92        & 0.86    & 0.81  \\
MSMARCO                  & 0.83   & 0.81   & 0.75        & 0.78        & 0.76    & 0.73  \\
\bottomrule
\end{tabular}
}
\end{table*}
\begin{table*}[h]
\centering
\caption{Transferability across Retrievers.}
\label{tab:impact-retriever-llm}
\resizebox{0.65\linewidth}{!}{
\begin{tabular}{c|ccc|ccc}
\toprule
\multirow{2}{*}{Retriever} & \multicolumn{3}{c}{NQ} & \multicolumn{3}{c}{MSMARCO} \\
\cmidrule(lr){2-4}
\cmidrule(lr){5-7}
                           & F1$\uparrow$    & L-ASR$\uparrow$  & V-ASR$\uparrow$ & F1$\uparrow$      & L-ASR$\uparrow$   & V-ASR$\uparrow$   \\
\midrule
Contriever                 & 0.89  & 0.90   & 0.85  & 0.70    & 0.83    & 0.81    \\
Contriever-ms              & 0.79  & 0.82   & 0.80  & 0.60    & 0.64    & 0.66    \\
ANCE                       & 0.63  & 0.70   & 0.68  & 0.47    & 0.56    & 0.60    \\
\bottomrule
\end{tabular}
}
\end{table*}
\begin{figure*}[ht] 
  \centering
  \includegraphics[width=\linewidth]{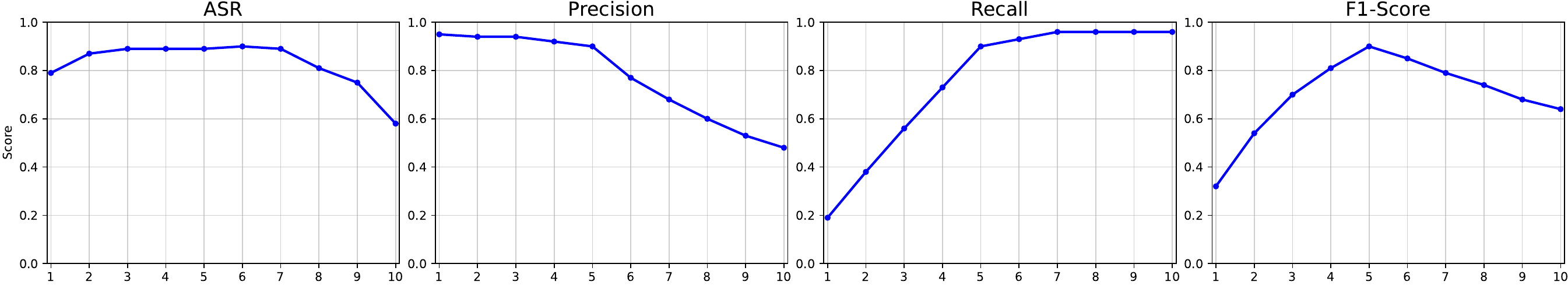} 
  \caption{Impact of Retrieval Scope ($k$) on NQ.}
  \label{fig:impact-of-k}
\end{figure*}

\subsection{Comparison with baselines}

The comprehensive results presented in \autoref{tab:attack_comparison} reveal that RefineRAG achieves a superior balance between effectiveness and stealthiness compared to all baselines. While PoisonedRAG (W) achieves a high retrieval F1-Score, its gradient-driven approach results in an anomalously high PPL (372.93) and frequent GE (6.54), making it easily detectable. 
Conversely, PoisonedRAG (B) addresses fluency by copying the query, but this heuristic leads to maximal RL with 1.00 and a high RR with 0.28, exposing it to deduplication filters. 
The Prompt Injection Attack similarly suffers from maximal redundancy (RR 1.00) due to its fixed template structure, while the Corpus Poisoning Attack fails completely in generation tasks with an ASR of 0.00. 
In contrast, RefineRAG secures the highest ASR on both NQ (0.90) and MSMARCO (0.83) while maintaining natural fluency (PPL 118) and registering the lowest GE and RR across all methods.

\subsection{Transferability Analysis}
We further evaluate the robustness of RefineRAG across different victim systems. 

\noindent\textbf{Transferability across Victim LLMs.}
As shown in \autoref{tab:impact-of-llms}, when tested against six diverse open-source LLMs, namely Llama2-7B (Llama2), Vicuna-7B (Vicuna), DeepSeek-R1, DeepSeek-V3, Qwen2.5-7B (Qwen2.5) and Qwen3-Max (Qwen3), RefineRAG maintains consistently high ASR scores ranging from 0.81 to 0.92 on the NQ dataset. This demonstrates strong model-agnostic transferability driven by its semantic optimization. 

\noindent\textbf{Transferability across Retrievers.}
In terms of retriever generalization, \autoref{tab:impact-retriever-llm} indicates that adversarial samples optimized on Contriever transfer effectively to unseen retrievers such as Contriever-msmarco (Contriever-ms) and ANCE. Although the F1-score decreases due to domain shifts, the attack maintains a significant ASR of up to 0.70 on NQ dataset, indicating that the generated texts occupy a broad toxic region in the embedding space. 

\noindent\textbf{Impact of Retrieval Scope.}
For the retrieval scope $k$ shown in \autoref{fig:impact-of-k}, we observe a non-linear relationship where attack performance peaks at a scope of 5 and gradually declines as $k$ increases to 10. This trend confirms the dilution effect where an excessive number of retrieved benign documents weakens the adversarial context, a phenomenon consistent with prior findings in the field.

\subsection{Ablation Study}

\begin{table*}[t]
\centering
\caption{Analysis of the contribution of each stage in RefineRAG.}
\label{tab:ablation_study_components}
\resizebox{0.75\linewidth}{!}{
\begin{tabular}{c|ccc|ccc}
\toprule
\multirow{2}{*}{Attack Method} & \multicolumn{3}{c}{NQ}                        & \multicolumn{3}{c}{MSMARCO}                   \\
\cmidrule(lr){2-4}
\cmidrule(lr){5-7}
                               & F1$\uparrow$            & L-ASR$\uparrow$         & V-ASR$\uparrow$         & F1$\uparrow$            & L-ASR$\uparrow$         & V-ASR$\uparrow$         \\
\midrule
No-I                    & 0.63          & 0.72          & 0.69          & 0.53          & 0.71          & 0.75          \\
No-II                   & 0.73          & 0.86          & 0.82          & 0.54          & 0.75          & 0.79          \\
RefineRAG                      & \textbf{0.89} & \textbf{0.90} & \textbf{0.85} & \textbf{0.70} & \textbf{0.83} & \textbf{0.81} \\
\bottomrule
\end{tabular}
}
\end{table*}
To verify the necessity of our two-stage design, we conduct ablation experiments by systematically removing key components of the framework. 
We first evaluate the configuration designated as No-I where Stage I is removed and WLO is applied directly to initial texts. This modification causes a sharp performance drop on the NQ dataset, with the L-ASR falling from 0.90 to 0.72 and the V-ASR decreasing from 0.85 to 0.69. The F1-score similarly declines from 0.89 to 0.63. We also observe a significant performance reduction on the MSMARCO dataset, confirming that micro-refinement relies heavily on a high-quality semantic foundation. 
Conversely, removing Stage II and using only the output from Stage I, a setting named No-II, results in a substantial decrease in retrieval effectiveness. Specifically, the retrieval F1-score drops to 0.73 compared to the 0.89 achieved by the full model, leading to a corresponding decline in ASR. These results demonstrate that the synergy between macro-level toxicity generation and micro-level retrieval optimization is essential for the framework’s overall success.

\subsection{Parameter Analysis}
\label{sec:sensitivity_analysis}
\begin{table*}[t]
\centering
\caption{Impact of Attacker Generators.}
\label{tab:attacker_llm}
\resizebox{0.7\linewidth}{!}{
\begin{tabular}{c|ccc|ccc}
\toprule
\multirow{2}{*}{Attack Model} & \multicolumn{3}{c}{NQ} & \multicolumn{3}{c}{MSMARCO} \\
\cmidrule(lr){2-4}
\cmidrule(lr){5-7}
                              & F1$\uparrow$    & L-ASR$\uparrow$  & V-ASR$\uparrow$ & F1$\uparrow$      & L-ASR$\uparrow$   & V-ASR$\uparrow$   \\
\midrule
Deepseek-V3                   & 0.89  & 0.90   & 0.85  & 0.70    & 0.83    & 0.81    \\
Qwen3-Max                         & 0.91  & 0.85   & 0.82  & 0.74    & 0.83    & 0.83    \\
\bottomrule
\end{tabular}
}
\end{table*}
\begin{table*}[ht]
\centering
\caption{Impact of Stage I Iterations.}
\label{tab:refinerag_rounds}
\resizebox{0.65\linewidth}{!}{
\begin{tabular}{c|ccc|ccc}
\toprule
\multirow{2}{*}{$T$-Value} & \multicolumn{3}{c}{NQ} & \multicolumn{3}{c}{MSMARCO} \\
\cmidrule(lr){2-4}
\cmidrule(lr){5-7}
                              & F1$\uparrow$    & L-ASR$\uparrow$  & V-ASR$\uparrow$ & F1$\uparrow$      & L-ASR$\uparrow$   & V-ASR$\uparrow$   \\
\midrule
3                       & 0.84          & 0.84          & 0.79          & 0.63          & 0.72          & 0.73          \\
4                       & \textbf{0.89} & \textbf{0.90} & \textbf{0.85} & \textbf{0.70} & \textbf{0.83} & \textbf{0.81} \\
5                       & 0.88          & 0.86          & 0.83          & 0.67          & 0.79          & 0.72    \\
\bottomrule
\end{tabular}
}
\end{table*}

\noindent\textbf{Robustness to Generator LLM Choice.}
To determine whether RefineRAG depends on a specific generative architecture, we fix the victim models and compare performance using two different generators in Stage I: DeepSeek-V3 and Qwen3-Max. 
The results demonstrate that the framework is highly robust to the choice of the attacker's generator, with final performance metrics remaining nearly identical across both models. For instance, on the NQ dataset, the L-ASR is 0.90 for DeepSeek-V3 compared to 0.85 for Qwen3-Max, while on MSMARCO, the performance is identical at 0.83. These findings confirm that the attack effectiveness is driven by the framework's optimization strategy rather than the inherent capability of a specific generator.

\label{sec:sensitivity_refinerag_hyperparams} 

\noindent\textbf{Optimal Rounds $T$ for Macro-Generation.}
We evaluate the impact of the number of iteration rounds $T$ in Stage I by testing values from the set $\{3, 4, 5\}$. Performance peaks at $T=4$, where RefineRAG achieves an F1-score of 0.89 on NQ and 0.70 on MSMARCO, alongside the highest ASR values. 
With fewer rounds ($T=3$), the mixed strategy underperforms due to insufficient convergence, yielding lower ASR scores. 
Conversely, increasing the rounds to $T=$5 leads to a slight degradation in performance, likely due to noise introduced by excessive iterations. Consequently, we adopt $T=4$ as the default setting to balance generation quality and stability.

\begin{table*}[ht]
\centering
\caption{Impact of Stage II WLO Iterations.}
\label{tab:refinerag_maxiter_sensitivity}
\resizebox{0.65\linewidth}{!}{
\begin{tabular}{c|ccc|ccc}
\toprule
\multirow{2}{*}{Iterations} & \multicolumn{3}{c}{NQ} & \multicolumn{3}{c}{MSMARCO} \\
\cmidrule(lr){2-4}
\cmidrule(lr){5-7}
                              & F1$\uparrow$    & L-ASR$\uparrow$  & V-ASR$\uparrow$ & F1$\uparrow$      & L-ASR$\uparrow$   & V-ASR$\uparrow$   \\
\midrule
5                       & 0.85          & \textbf{0.91} & \textbf{0.87} & 0.65          & 0.79          & 0.80          \\
10                      & \textbf{0.89} & 0.90          & 0.85          & 0.70          & \textbf{0.83} & \textbf{0.81} \\
15                      & \textbf{0.89} & 0.88          & 0.84          & \textbf{0.74} & 0.81          & 0.79  \\
\bottomrule
\end{tabular}
}
\end{table*}
\begin{table*}[t]
\centering
\caption{Sensitivity to the number of candidate selections.}
\label{tab:refinerag_topk_ablation}
\resizebox{0.65\linewidth}{!}{
\begin{tabular}{c|ccc|ccc}
\toprule
\multirow{2}{*}{$K$-Value} & \multicolumn{3}{c}{NQ} & \multicolumn{3}{c}{MSMARCO} \\
\cmidrule(lr){2-4}
\cmidrule(lr){5-7}
                              & F1$\uparrow$    & L-ASR$\uparrow$  & V-ASR$\uparrow$ & F1$\uparrow$      & L-ASR$\uparrow$   & V-ASR$\uparrow$   \\
\midrule
10                       & 0.85          & 0.87          & \textbf{0.86} & 0.67          & 0.80          & 0.80          \\
20                       & \textbf{0.89} & \textbf{0.90} & 0.85          & 0.70          & \textbf{0.83} & \textbf{0.81} \\
30                       & \textbf{0.89} & 0.87          & 0.85          & \textbf{0.72} & 0.82          & 0.77   \\
\bottomrule
\end{tabular}
}
\end{table*}
\begin{table*}[ht]
\centering
\caption{Impact of Beam Size.}
\label{tab:refinerag_beam_size}
\resizebox{0.65\linewidth}{!}{
\begin{tabular}{c|ccc|ccc}
\toprule
\multirow{2}{*}{$B$-Value} & \multicolumn{3}{c}{NQ} & \multicolumn{3}{c}{MSMARCO} \\
\cmidrule(lr){2-4}
\cmidrule(lr){5-7}
                              & F1$\uparrow$    & L-ASR$\uparrow$  & V-ASR$\uparrow$ & F1$\uparrow$      & L-ASR$\uparrow$   & V-ASR$\uparrow$   \\
\midrule
1                        & 0.86          & 0.87          & 0.84          & 0.68          & 0.77          & 0.80          \\
3                        & \textbf{0.89} & \textbf{0.90} & 0.85          & 0.70          & \textbf{0.83} & \textbf{0.81} \\
5                        & 0.88          & 0.88          & \textbf{0.89} & \textbf{0.71} & 0.79          & 0.80    \\
\bottomrule
\end{tabular}
}
\end{table*}

\noindent\textbf{Sensitivity to WLO Iterations.}
We investigate the effect of the number of WLO iterations in Stage II by comparing performance at 5, 10, and 15 iterations. 
On the NQ dataset, while increasing iterations improves the retrieval F1-score, the ASR peaks at 5 iterations, suggesting that further refinement may weaken the adversarial signal. 
On MSMARCO, the ASR peaks at 10 iterations, even as the F1-score continues to rise. These trends indicate a trade-off where additional iterations enhance retrieval visibility but do not consistently improve the likelihood of misleading the LLM. 
We therefore select 10 iterations as the balanced configuration for our main experiment.


\noindent\textbf{Effect of MLM Candidate Count $k$.}
We analyze the sensitivity of Stage II to the number of replacement candidates predicted by the MLM, testing $K$ values of 10, 20, and 30. 
Increasing $K$ from 10 to 20 consistently improves performance. For example, the F1-score on NQ rises from 0.85 to 0.89, and the L-ASR on MSMARCO increases to 0.90. 
However, further increasing $K$ to 30 yields diminishing returns, with ASR decreasing on both datasets. 
This suggests that larger candidate sets may introduce semantic noise that dilutes the adversarial efficacy. 
Based on these results, we set $K$ to 20 as the default to optimize the balance between candidate diversity and attack precision.


\noindent\textbf{Influence of Beam Search Width $B$.}
Finally, we examine the impact of the beam size $B$ during the WLO process by comparing widths of 1, 3, and 5. 
Using a beam size of $B=3$ yields substantial improvements over the greedy approach ($B=1$), raising the L-ASR on MSMARCO from 0.77 to 0.83. 
Increasing the beam width further to $B=5$ provides only marginal gains in retrieval metrics and leads to lower ASR across most configurations. This indicates that larger beams may over-prioritize retrieval metrics at the expense of the adversarial signal. 
Therefore, we adopt $B=3$ as the default setting to achieve the best trade-off between attack success and computational efficiency.

\section{Discussion}

\subsection{Ethical Considerations}
Our primary motivation is to uncover RAG vulnerabilities in high-stakes domains before malicious exploitation. We strictly adhere to responsible AI principles. Experiments were conducted in an isolated simulation environment using public datasets. While commercial LLM APIs were utilized for simulation, the evaluation was strictly confined to our local setup, ensuring no impact on real-world systems. We aim to urge the prioritization of sanitization-based defenses against such stealthy threats.

\subsection{Limitations and Future Work}
We acknowledge three limitations that direct future research:
\begin{itemize}[leftmargin=*]
    \item \textbf{Computational Overhead:} The iterative MLM-based optimization in Stage II incurs higher computational costs than simple concatenation methods. Future work could explore distillation techniques to accelerate the process.
    \item \textbf{Dependency on Proxy Retrievers:} Our black-box transferability assumes embedding similarity between proxy and victim retrievers. Efficacy against radically different architectures (sparse retrievers) requires further investigation into ``universal'' perturbations.
    \item \textbf{Defense Evasion Boundaries:} While RefineRAG bypasses fluency-based filters, its robustness against advanced semantic defenses (external fact-checking) remains to be evaluated in future studies.
\end{itemize}

\section{Conclusion}
In this paper, we address the limitation of existing RAG poisoning attacks where effectiveness comes at the cost of stealthiness. By shifting from coarse-grained splicing to RefineRAG's fine-grained, word-level refinement, we synthesize poisoning texts that are both highly retrievable and linguistically coherent. Our experiments confirm that RefineRAG significantly outperforms SOTA methods in both success rates and stealthiness metrics. 
Moreover, our findings reveal a concerning level of transferability, where adversarial samples generated locally on proxy models can effectively compromise unknown, black-box RAG systems. This work underscores the urgent need for more sophisticated defense mechanisms capable of detecting fine-grained semantic perturbations, as traditional filters based on perplexity or repetition are insufficient against this new class of stealthy attacks. 

\bibliographystyle{splncs04}
\bibliography{main}

\begin{thebibliography}{10}
\providecommand{\url}[1]{\texttt{#1}}
\providecommand{\urlprefix}{URL }
\providecommand{\doi}[1]{https://doi.org/#1}

\bibitem{alkhalaf2024applying}
Alkhalaf, M., Yu, P., Yin, M., Deng, C.: Applying generative ai with retrieval augmented generation to summarize and extract key clinical information from electronic health records. Journal of biomedical informatics  \textbf{156},  104662 (2024)

\bibitem{bajaj2016msmarco}
Bajaj, P., Campos, D., Craswell, N., Deng, L., Gao, J., Liu, X., Majumder, R., McNamara, A., Mitra, B., Nguyen, T., Wang, S., Wang, X.: Ms marco: A human generated dataset for research on machine reading comprehension and question answering (2016), \url{https://arxiv.org/abs/1611.09268}

\bibitem{carlini2023poisoning}
Carlini, N., Tramer, F., Wallace, E., Jagielski, M., Herbert-Voss, A., Lee, K., Roberts, A., Brown, T., Song, D., Erlingsson, U., Oprea, A., Raffel, C.: Poisoning web-scale training datasets are easier than you might think. In: Proceedings of the IEEE Symposium on Security and Privacy (S\&P). pp. 1369--1387 (2023). \doi{10.1109/SP49137.2023.10179267}, \url{https://ieeexplore.ieee.org/document/10179267}

\bibitem{chen2024benchmarking}
Chen, J., Lin, H., Han, X., Sun, L.: Benchmarking large language models in retrieval-augmented generation. In: Proceedings of the AAAI Conference on Artificial Intelligence. vol.~38, pp. 16715--16723 (2024), \url{https://arxiv.org/abs/2311.16109}

\bibitem{chiang2023vicuna}
Chiang, W.L., Li, Z., Lin, Z., Sheng, Y., Wu, Z., Zhang, H., Zheng, L., Zhuang, S., Zhuang, Y., Gonzalez, J.E., Stoica, I., Xing, E.P.: Vicuna: An open-source chatbot impressing gpt-4 with 90\%* chatgpt quality (2023), \url{https://vicuna.lmsys.org/}

\bibitem{deepseek2024llm}
{DeepSeek-AI}: {DeepSeek LLM}: Scaling open-source language models with reinforcement learning (2024), \url{https://arxiv.org/abs/2401.02954}

\bibitem{devlin2019bert}
Devlin, J., Chang, M.W., Lee, K., Toutanova, K.: {BERT}: Pre-training of deep bidirectional transformers for language understanding. In: Proceedings of the 2019 Conference of the North American Chapter of the Association for Computational Linguistics: Human Language Technologies, Volume 1 (Long and Short Papers). pp. 4171--4186. Association for Computational Linguistics (2019). \doi{10.18653/v1/N19-1423}, \url{https://aclanthology.org/N19-1423}

\bibitem{ebrahimi2018hotflip}
Ebrahimi, J., Rao, A., Lowd, D., Dou, D.: Hotflip: White-box adversarial examples for text classification. In: Proceedings of the 56th Annual Meeting of the Association for Computational Linguistics (Volume 2: Short Papers). pp. 382--387. Association for Computational Linguistics (2018). \doi{10.18653/v1/P18-2061}, \url{https://aclanthology.org/P18-2061}

\bibitem{edge2024graphrag}
Edge, D., Trinh, H., Cheng, N., Bradley, J., Chao, A., Mody, A., Truitt, S., Metropolitansky, D., Ness, R.O., Larson, J.: From local to global: A graph {RAG} approach to query-focused summarization (2024), \url{https://arxiv.org/abs/2404.16130}

\bibitem{honnibal2020spacy}
Honnibal, M., Montani, I., Van~Landeghem, S., Boyd, A., et~al.: spacy: Industrial-strength natural language processing in python  (2020)

\bibitem{huang2023catastrophic}
Huang, Y., Gupta, S., Xia, M., Li, K., Chen, D.: Catastrophic jailbreak of open-source llms via exploiting generation (2023), \url{https://arxiv.org/abs/2310.06987}

\bibitem{izacard2022contriever}
Izacard, G., Caron, M., Hosseini, L., Riedel, S., Bojanowski, P., Joulin, A., Grave, E.: Contriever: Improving contrastive learning for unsupervised text retrieval. In: Proceedings of the 39th International Conference on Machine Learning (ICML). Proceedings of Machine Learning Research, vol.~162, pp. 9745--9758. PMLR (2022), \url{https://proceedings.mlr.press/v162/izacard22a.html}

\bibitem{izacard2022unsupervised}
Izacard, G., Caron, M., Hosseini, L., Riedel, S., Bojanowski, P., Joulin, A., Grave, E.: Unsupervised dense information retrieval with contrastive learning. Transactions on Machine Learning Research  (2022), \url{https://openreview.net/forum?id=kXwdL1cWO5}

\bibitem{jin2020textfooler}
Jin, D., Jin, Z., Zhou, J.T., Szolovits, P.: Is {BERT} really robust? a strong baseline for natural language attack on text classification and entailment. In: Proceedings of the AAAI Conference on Artificial Intelligence. vol.~34, pp. 8018--8025 (2020). \doi{10.1609/aaai.v34i05.6304}, \url{https://ojs.aaai.org/index.php/AAAI/article/view/6304}

\bibitem{kwiatkowski2019natural}
Kwiatkowski, T., Palomaki, J., Redfield, O., Collins, M., Parikh, A., Alberti, C., Epstein, D., Polosukhin, I., Devlin, J., Lee, K.: Natural questions: A benchmark for question answering research. Transactions of the Association for Computational Linguistics  \textbf{7},  453--466 (2019). \doi{10.1162/tacl_a_00276}, \url{https://aclanthology.org/Q19-1026}

\bibitem{NEURIPS2020_6b493230}
Lewis, P., Perez, E., Piktus, A., Petroni, F., Karpukhin, V., Goyal, N., K\"{u}ttler, H., Lewis, M., Yih, W.t., Rockt\"{a}schel, T., Riedel, S., Kiela, D.: Retrieval-augmented generation for knowledge-intensive nlp tasks. In: Larochelle, H., Ranzato, M., Hadsell, R., Balcan, M., Lin, H. (eds.) Advances in Neural Information Processing Systems. vol.~33, pp. 9459--9474. Curran Associates, Inc. (2020), \url{https://proceedings.neurips.cc/paper_files/paper/2020/file/6b493230205f780e1bc26945df7481e5-Paper.pdf}

\bibitem{li2025cparag}
Li, C., Zhang, J., Cheng, A., Ma, Z., Li, X., Ma, J.: Cpa-rag: Covert poisoning attacks on retrieval-augmented generation in large language models (2025), \url{https://arxiv.org/abs/2505.19864}

\bibitem{li2020bertattackadversarialattackbert}
Li, L., Ma, R., Guo, Q., Xue, X., Qiu, X.: Bert-attack: Adversarial attack against bert using bert (2020), \url{https://arxiv.org/abs/2004.09984}

\bibitem{perez2022ignore}
Perez, F., Ribeiro, I.: Ignore previous prompt: Attack techniques for language models (2022), \url{https://arxiv.org/abs/2211.09527}

\bibitem{radford2019language}
Radford, A., Wu, J., Child, R., Luan, D., Amodei, D., Sutskever, I., et~al.: Language models are unsupervised multitask learners. OpenAI blog  \textbf{1}(8), ~9 (2019)

\bibitem{rizqullah2023qasina}
Rizqullah, M.R., Purwarianti, A., Aji, A.F.: Qasina: Religious domain question answering using sirah nabawiyah (2023), \url{https://arxiv.org/abs/2310.08102}

\bibitem{salemi2024evaluating}
Salemi, A., Zamani, H.: Evaluating retrieval quality in retrieval-augmented generation. In: Proceedings of the 47th International ACM SIGIR Conference on Research and Development in Information Retrieval. pp. 2185--2189 (2024). \doi{10.1145/3626772.3657754}, \url{https://dl.acm.org/doi/abs/10.1145/3626772.3657754}

\bibitem{touvron2023llama2}
Touvron, H., Martin, L., Stone, K., Albert, P., Almahairi, A., Babaei, Y., Bashlykov, N., Batra, S., Bhargava, P., Bhosale, S., et~al.: Llama 2: Open foundation and fine-tuned chat models (2023), \url{https://arxiv.org/abs/2307.09288}

\bibitem{wang2024metmap}
Wang, G., Li, Y., Liu, Y., Deng, G., Li, T., Xu, G., Liu, Y., Wang, H., Wang, K.: Metmap: Metamorphic testing for detecting false vector matching problems in llm augmented generation. In: Proceedings of the 2024 IEEE/ACM First International Conference on AI Foundation Models and Software Engineering (FORGE). pp. 12--23 (2024). \doi{10.1145/3650105.3652297}

\bibitem{wei2023jailbroken}
Wei, A., Haghtalab, N., Steinhardt, J.: Jailbroken: How does llm safety training fail? (2023), \url{https://arxiv.org/abs/2307.02483}

\bibitem{xiong2021approximate}
Xiong, L., Xiong, C., Li, Y., Tang, K.F., Liu, J., Bennett, P.N., Ahmed, J., Overwijk, A.: Approximate nearest neighbor negative contrastive learning for dense text retrieval. In: International Conference on Learning Representations (ICLR) (2021), \url{https://openreview.net/forum?id=zeFrfgyZln}

\bibitem{yepes2024financial}
Yepes, A.J., You, Y., Milczek, J., Laverde, S., Li, R.: Financial report chunking for effective retrieval augmented generation (2024), \url{https://arxiv.org/abs/2402.05131}

\bibitem{zhang2023enhancing}
Zhang, B., Yang, H., Zhou, T., Babar, M.A., Liu, X.Y.: Enhancing financial large language models with retrieval-augmented generation (2023), \url{https://arxiv.org/abs/2308.14081}

\bibitem{zhao2023medrag}
Zhao, X., Liu, S., Yang, S.Y., Miao, C.: Medrag: Improving medical diagnosis with retrieval-augmented generation (2023), \url{https://arxiv.org/abs/2306.02322}

\bibitem{zhong2024poisoning}
Zhong, Z., Huang, Z., Wettig, A., Chen, D.: Poisoning retrieval corpora: How to mislead retrieval-augmented generation. In: International Conference on Learning Representations (ICLR) (2024), \url{https://openreview.net/forum?id=1EB1fSj23k}

\bibitem{zou2024poisonedrag}
Zou, W., Geng, R., Wang, B., Jia, J.: Poisonedrag: Knowledge corruption attacks to retrieval-augmented generation of large language models (2024), \url{https://arxiv.org/abs/2402.07867}

\end{thebibliography}
\end{document}